# Feedback patterns in simulating intestinal wall motions: interdisciplinary approach to the motility mechanisms.

Garri Davydyan

Appletree Medical Group

email address: garri.davydyan@gmail.com

ABSTRACT

Ability of smooth muscles to contract in response to distension plays a crucial role in motor function of intestine. Qualitative analysis of dynamical models using myogenic active property of smooth muscles has shown well agreement with physiologic data. Considered as a self-regulatory unit, function of gastrointestinal (GI) segment is assumed to be regulated by integration of basis patterns providing accumulation and propagation of intestinal content. By implementing external, depending on neural system, variable to the previous model, and considering two attaches to one another reservoirs as a physical analogue of the segmental partition of intestine, a system of six ODE equations, three for each reservoir, describes coordinated wall motions and propagation of the content from one reservoir to another. It was shown that besides negative feedback (NFB), other functional patterns, namely positive feedback (PFB) and reciprocal links (RL) are involved in regulations of filling-emptying cycle. Being integrated in a whole functional system these three patterns expressed in a matrix form represent basis elements of imaginary part of coquaternion which with identity basis component is an algebraically closed structure under addition and multiplication of its elements. A coquaternion ring may be considered as a model of inner self-regulatory functional structure providing coordinated wall motions of GI tract portions.

KEY WORDS: GI motility, smooth muscle tone, NFB, PFB, RL, coquaternion.

INTRODUCTION

Motility of GI tract is regulated by numerous factors of which myogenic active property of smooth muscles of the wall to create contractile forces when the stretching reaches a threshold, is a key feature [1-10].
In a previous work [1] we assumed that the stress arising in intestinal wall during its distension is a sum of passive, depending purely on mechanical properties of the wall, and active, determined by the ability of smooth muscles to contract in response to distension, components [11-14]. Also considering a smooth muscle tone ($N$) and a circular distension of the wall ($\varepsilon$) of a distensible reservoir as variables and in a given conditions of opening and closing valves at the ends of the reservoir a mathematical model of the wall motions of a single intestinal segment was formulated (*), (Fig. 1):

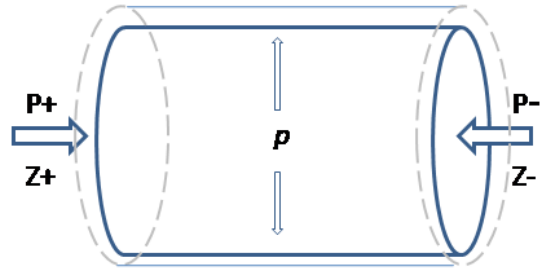

**Figure 1** Distensible cylindrical reservoir of flexible radius R and constant length L as a prototype of intestinal segment whose circular smooth muscles are endowed with myogenic active properties

$$\frac{d\varepsilon}{dt} = \frac{1}{2V_0} \left( \frac{P_+}{Z_+} + \frac{P_-}{Z_-} - \frac{E\varepsilon + N}{2\varepsilon + 1} \frac{h_0}{R_0} \frac{1}{Z} \right)$$

$$\frac{dN}{dt} = -k_3 N^3 + k_2 N^2 - k_1 N + \varkappa\varepsilon + k_0(t)$$

(*)

Here, $P_+$, $P_-$ and $Z_+$, $Z_-$ are pressures and resistances at the ends of the reservoir, respectively; $V_0$ - reservoir volume corresponding to the initial radius $R_0$; $k_i$, $\varkappa$ - coefficients.

The first equation is based on the mass conservation law; the second one reflects the dependence of smooth muscle tone on the flow of ionized $Ca^{2+}$ through the membranes of smooth muscle cells.

Despite quite general assumptions, the results were in a good agreement with physiologic data. The character of phase trajectories clearly demonstrated that contractile activity of intestine could be initiated at lesser threshold level, if the smooth muscles of intestinal wall initially had high tone and vice versa. Obtained results elucidated physiological mechanisms of the well-known Bayliss- Starling law [13, 15].

Another way to understand motility mechanisms is to use general principles of functional organization of biologic systems, i.e. system's approach to the motility of GI segments considered as autonomous self-regulatory units.

Some authors, describing physiologic functions of GI tract mentioned feedback as possible regulatory mechanism of motility [16, 17]. The goal of this work is to find out which feedback elements exist as patterns of the functional system and how they are involved in motility regulations.

1. FORMULATION OF THE MODEL

Because of general and quite demonstrative features of the system (*), it was taken as a basic model for further consideration in this work.

For the purposes of this work some of the previous statements [1] will be clarified. Smooth muscles of intestinal wall are organized in a two-layer coat, - circular and longitudinal ones, but only the circular layer has been considered to provide the wall movements. So, the length of the cylinder remained constant during its filling and emptying. The end parts of the tube contained valves with resistances $Z_+$ and $Z_-$, so that the differences in the intra-luminal pressure created the conditions for opening and closing the valves, thus causing the content to move into the reservoir, fill it until threshold level is reached and be evacuated due to active contractions of the wall.

Conditions for opening and closing the valves were formulated as follows:

$$P^+ - p > p_+^* , \quad p - P^- > p_-^* ,$$

where $P^+$ and $P^-$ are pressures at the proximal (input) and distal (output) ends of the reservoir, respectively, $p$ is a pressure inside the reservoir, $p_+^*$ and $p_-^*$ are threshold pressure differences causing closing input and output valves, respectively.

The first inequality which is related to the conditions of the input valve shows that in order to open the input valve and begin filling the input pressure $P^+$ should exceed the pressure inside the reservoir on a threshold level $p_+^*$. When the pressure inside the reservoir becomes "high" enough, and the condition does not satisfy the first inequality, the input valve closes and the reservoir enters into the isovolumetric stage when the muscle tone and reservoir's pressure continue to increase due to autocatalytic process providing an increase in a muscle tone. The pressure changes inside the reservoir permits both possibilities for the input valve. Closing the input valve prevents retrograde emptying of the content in normal physiologic conditions.

The second inequality shows that in order to open the output valve the pressure inside the reservoir must exceed the pressure into the outlet tube plus the threshold level $p_-^*$. If the pressure inside the reservoir is not high enough to open the outlet valve, but reaches the level when input valve closes, the both valve are closed and the conditions of the reservoir undergo isovolumetric states. Isovolumetric rise of the intraluminal pressure is an autocatalytic process resulted in the accumulation of the energy to increase contractile forces to provide efficacy of emptying.

For simplicity purposes conditions for the valves were formulated in a way that permits any combination except opening both valves. Input and output pressures and the threshold pressures also were given as the constants. Pressure outside the reservoir were not established as functions of inlet and outlet tubes, because only a single cylinder (tube segment) with myogenic properties has been considered in the previous model. For other details see [1].

According to the previous model (*) when the pressure in the inlet tube exceeds the pressure inside the reservoir the input valve opens and filling begins. During the filling (Fig. 2) increase both in the tone $N$ and distension $\varepsilon$ of a circular muscular layer cause increase in the pressure inside the reservoir. After reaching a threshold stress-strained level of the reservoir's wall (intersection with $d\varepsilon/dt = 0$ isocline), active contractions of circular muscles begin. Increase in the pressure inside the reservoir opens the input valve which causes retrograde emptying the content into the inlet tube. This process (filling and retrograde emptying) may have undulating character until the system reaches an equilibrium state. This is a general scenario, determined by the conditions of opening and closing the valves, when the pressure inside the reservoir does not reach the level when the output valve opens. If the stress-strained conditions dynamics predisposes opening the output valve, normally it will follow isovolumetric increase in the luminal reservoir pressure when both valves closed. Opening of the output valve, while the input one closed, causes movement of the content into the outlet tube and possibly to another, distally located, segment. On the phase plane portions of the trajectories above isobar $\pi^-$ indicate the reservoir emptying which is shown initially as decreasing the muscle stretching and slightly increasing the muscle tone and then, after intersecting with the $\frac{dN}{dt} = 0$ isocline, decreasing the tone coincidentally with decreasing the stretching. If the reservoir pressure is high enough, contractions of the reservoir walls during emptying are accompanied by decrease in the muscle tone without the initial phase of its temporary increase (upper parts of the phase portrait corresponding to high tone and stretching).

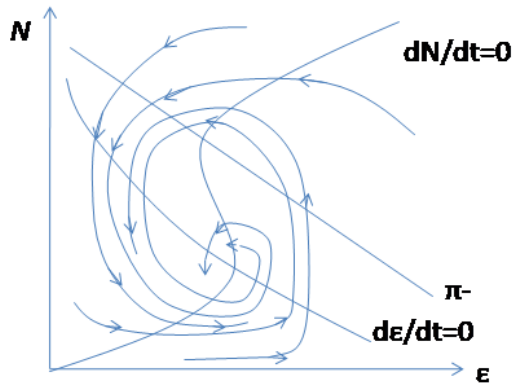

**Figure 2** Sketch of phase trajectories of stress-strain conditions of distensible cylindrical reservoir of flexible radius R and constant length L. Filling, isovolumetric and emptying areas on the plane are shown as corresponding portions of the curves.

For simplicity some isobars are omitted. Between isocline dε/dt=0 and isobar π- is an area of isovolumetric increase in pressure.

As a separable morphological unit an intestinal segment and/or even the whole intestine represent a relatively closed autonomous functional structure, - a biologic system (subsystem), which must possess self -regulatory properties (mechanisms) providing the system with morphological steadiness and functional stability. There are numerous data confirming the existence of NFB, PFB and RL feedback regulatory loops on different functional levels: between organs, inside biological cells and as components of biochemical reactions [18-20]. Matrices used in linearized dynamical models representing these basis functional patterns are:

$$S_0 = \begin{pmatrix} 0 & + \\ - & 0 \end{pmatrix}, \quad S_1 = \begin{pmatrix} + & 0 \\ 0 & - \end{pmatrix}, \quad S_2 = \begin{pmatrix} 0 & + \\ + & 0 \end{pmatrix}.$$

$S_0$, $S_1$, $S_2$ are matrices of NFB, PFB and RL, respectively (Fig. 3) .

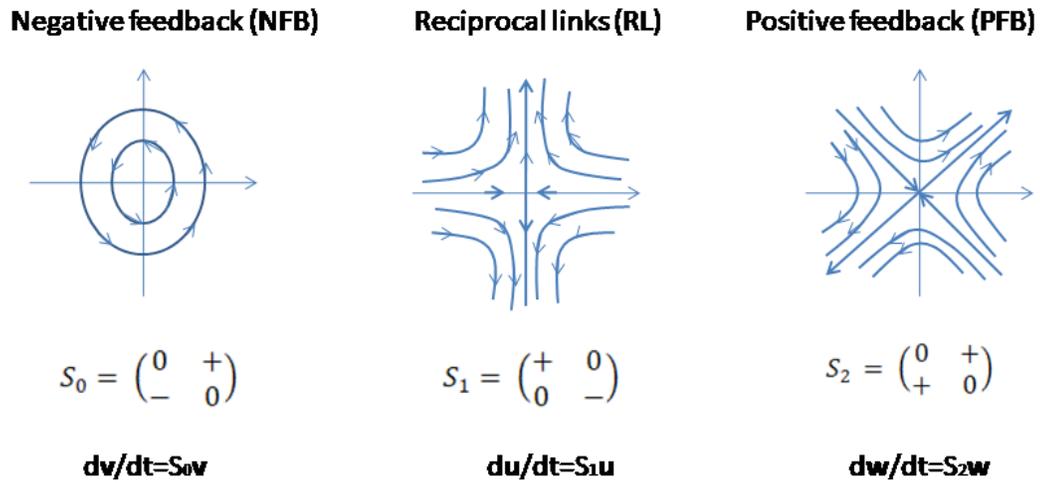

Figure 3   Sketch of dynamical images of NFB, PFB and RL patterns obtained from ODE. Matrices of NFB, PFB and RL are operators of ODE. Variables are expressed in a vector form.

Because of special functional and mathematical properties of these patterns [21-24], it was proposed that they may serve as a universal functional basis of steady morphological elements like molecules, cells, organs, etc., or, in other words, form internal functional elements of biologic systems [25-27]. Finding of basis functional patterns regulating motility function of GI segment will further validate the correctness of obtained modeling features as well as to help in understanding the invariant functional mechanisms regulating GI motility.

## 1.1 FEEDBACK FUNCTIONAL PATTERNS IN MOTILITY FUNCTION

Despite possible differences in scenarios determined by the parameters values and conditions of the valves, behavior of the system (*) around the equilibrium point always corresponds to *steady focus*. It can be demonstrated after linearization of the system around equilibrium point and transforming it to the Jourdan form.

The flow of linearized system (*) of two variables ($N$, $\varepsilon$) corresponds to *steady focus* and determined by a Jourdan matrix $A = \begin{pmatrix} -a & -c \\ +c & -a \end{pmatrix}$ whose entries are real numbers. Considered $A$ as a sum of two matrices $B = \begin{pmatrix} -a & 0 \\ 0 & -a \end{pmatrix}$ and $C = \begin{pmatrix} 0 & -c \\ +c & 0 \end{pmatrix}$, it is easy to see that the regulatory system related to $A$ contains NFB regulatory pattern between $N$ and $\varepsilon$ variables which is represented by matrix $C$ and its phase imaging - *center*. Matrix $C$ is fused functionally with the matrix $B$ in a steady self-regulatory functional structure or a pattern described by steady focus. Matrix $B$ represents *steady node*. (Fig. 4). Despite unsteadiness of $C$ considered as a separable functional structure [28], it always exists as a component of $A$.

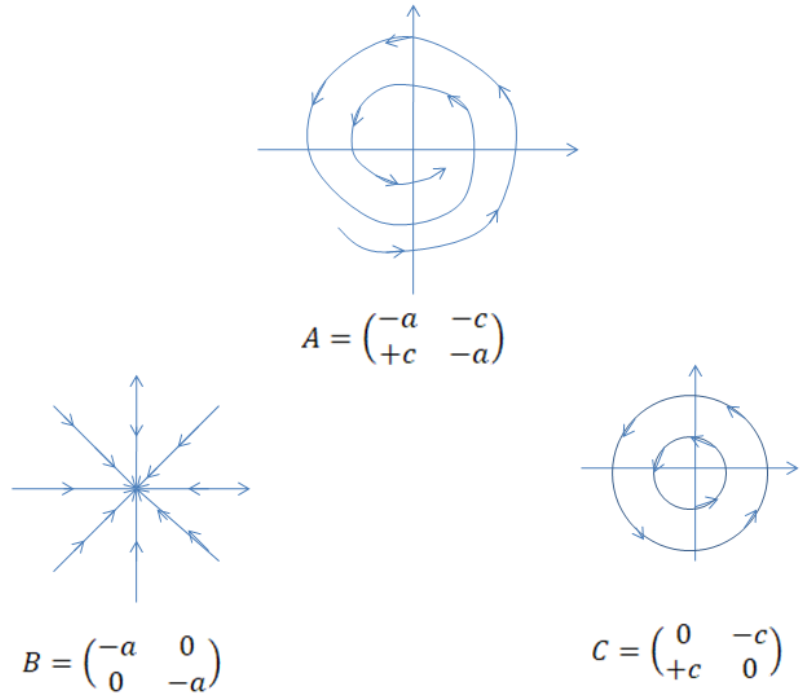

Figure 4  Matrix A representing *steady focus* can be presented as a sum of *steady node* matrix (B) and a matrix representing *center* (C). Center is an image of NFB. Steady node is an image of identity matrix multiplied by -α

Muscle tone $N$ and deformations $\varepsilon$ collaborate as elements of a hierarchical system in which an active stress is a dominating regulatory element determining the range of muscles stretching when distension forces are applied. The higher the initial value of the muscle tone, the less distension of the muscle is needed to increase the pressure inside the reservoir and vice versa. Physiologically a NFB component between $N$ and $\varepsilon$ provides optimal degree of the muscle distension depending on the initial values of the muscle tone. NFB "organizes" $N$ and $\varepsilon$ characters (variables) in a self-regulatory subsystem responsible for creation of the pressure inside the reservoir which is the main physical factor moving the intestinal content along GI tract. Reservoir emptying diminishes the luminal pressure thus returning it to the conditions when filling begins with the new filling-emptying cycle.

Filling and emptying trajectories of the first model can be regarded as two simultaneous processes taking place in two attached to one another reservoirs (Fig. 5).  In this case filling trajectories will correspond to the distal reservoir (II), while the emptying curves on the phase plane -to the evacuation from the proximal reservoir (I) (Fig. 6).

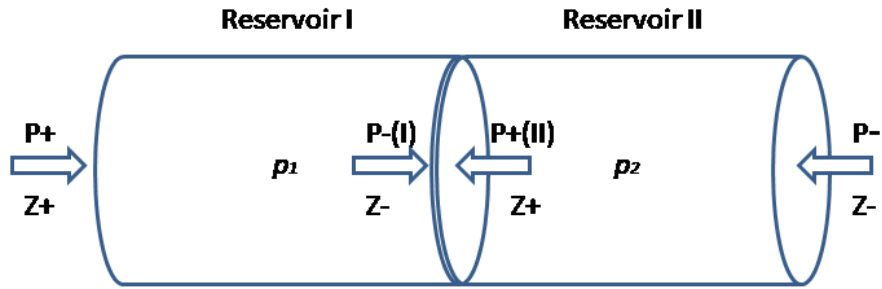

Figure 5  Cascade of two reservoirs. Conditions for opening and closing valves at the unction are reformulated so that the output and input threshold pressures are equalized to the luminal pressures
P-(I)=p2 , P+ (II)=p1

In order both processes, emptying of the reservoir (I) and filling of the reservoir (II), are being simultaneous resulting in the movement of the content in caudal direction, conditions for opening and closing output valve for reservoir (I) and input valve for reservoir (II) should be reformulated in a way to synchronize their work. Consider $P_1^- = p_2$ and $P_2^+ = p_1$ , and substitute them to the inequalities written separately for proximal and distal reservoirs

$$p_1 - P_1^- > p_-^* \ , \ P_2^+ - p_2 > p_+^* \quad .$$

Condition for "synchronous" opening of the valves at the reservoirs unction will satisfy $p_-^* = p_+^*$ .

For simplicity, the resistances of the reservoirs are accommodated to prevent pathological flow.

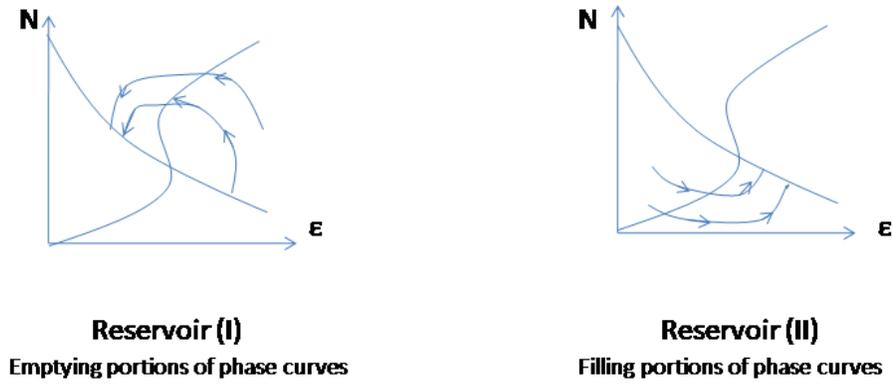

Figure 6  Sketch of emptying and filling trajectories of two connected reservoirs
Isovolumetric areas and the curves portions are not shown

Mechanism of propagation of intestinal content is determined by coordinated contractions and relaxations of anatomically or functionally separated segments along the GI tube. It is provided by creation of a pressure gradient between neighbor segments and coordination of their contractility functions.

It seems that mechanism coordinating motility functions of anatomically connected intestinal segments was developed phylogenetically by splitting of initially irreducible regulatory elements on anatomically and functionally reciprocal parts [3, 5]. In this sense two neighbor segments should have besides similar mechanisms for the wall motions some additional regulatory factors and (or) anatomical structures preventing anterograde or retrograde emptying of the content until it is fermented, partially absorbed and is ready to be evacuated. Not only anatomically GI tract parts are differ and separated from one another like small and large intestines, separated by ileum- cecal valve, etc., but additional, functional, mechanisms should also exist. For example, two chambers (segments) of small intestine can be temporarily separated from one another by circular muscle contractions creating additional resistance preventing the content from escape.

Not only internal (myogenic) mechanisms, but also external (neuronal) factors are involved in regulations of the smooth muscles contractility and intestinal wall motions. They also coordinate motility functions of neighbor segments. Among those factors are Auerbach (intermuscular) and Meysner (submucous) neuronal plexuses of intestine having direct activating and inhibiting actions on smooth muscles via neurotransmitter acetylcholine (Ach).

Functional availability of Ach is directly related to the Na-K pump mechanism. Its cyclic functional structure characterizes by phases of the passive trans-membrane flow of Na+ inside the cell and its active transport from the intracellular space backward against electrochemical gradient. Instead, K+ is transported inside the cells, and the both Na+ and K+ flows occur against electrochemical gradients.

Passive flow of Na through the cell's membrane eventually causes increase in its concentration inside the cell. When the amount of ionized Na inside the neuronal cell reaches a threshold, it causes opening of the additional channels for Na+ and massive flow of the cation inside the cell. This process leads to the generation of an electrical spike which propagates along the axon to the neuromuscular unction. The next phase after the conduction of electrical impulse is a refractory stage when excessive amount of intracellular Na+ is being transported from the cell into the extracellular space. This mechanism is well described by van der Pol harmonic oscillator and Lienare ODE where kinetics of Na+ is described by cubic equations [32].

Amount of Ach being released into neuromuscular unction is determined by electrical impulses (spikes) propagating along the axon. In this sense the spikes generating mechanism is equivalent to the changes in the external activator associated with Ach.

Therefore, the function of external activator should include areas where amount of the activator decreases the changes in its concentration and areas where concentration of the activator (Ach or ionized intracellular Na+) substantially increases its amount. The latter domain corresponds to the initiation of active muscle contractions or generation of an action potential [29-32]. Considering not only $Na^+$-dependent, but also other factors ($Ca^{2+}$ activation mechanisms associated with muscle tone) involved in regulations of external activator, hypothetical equation showing changes of the amount of the external activator should reflect the form of transmembrane kinetics of Na+ and its availability; it also should contain other factors, including the smooth muscle tone due to direct actions of Ach on extracellular Ca2+ responsible for smooth muscle contractility.

Hypothetical equation postulated for external activator $f$ is:

$$\frac{df}{dt} = -l_3 f^3 + l_2 f^2 - l_1 f + r(N) \qquad (1)$$

Here $f$ - external activator, $l_{1-3}$ coefficients, $r(N)$ - additional factors, including muscle tone affecting [influencing] changes in the external activator. Explicit form for $r(N)$ will be shown further in the text.

The simplest form describing the influence of external activation factor $f$ on myogenic activation of the smooth muscles should reflect its inhibitory actions on the changes of the smooth muscle tone $\frac{dN}{dt}$ during reservoir filling and positive, excitatory, stimulation while emptying.
It has a physiologic sense: during the reservoir filling distension of the reservoir wall is accompanied by the increase in the muscle tone and the pressure inside the reservoir; in order to alleviate movement of the content into the reservoir external factors should inhibit activation of muscle tone and slow down the keep growing intraluminal pressure to allow more content to enter the reservoir until input valve closes.

On the other hand, during emptying external activator should act in a way to stimulate myogenic activity of the smooth muscles in order to increase the contractile forces.

Thus, influence of neuronal activation factor $f$ on internal (myogenic) mechanisms should affect different stages, i.e. filling and emptying, in a reciprocal manner. When two intestinal segments are anatomical neighbors, external factors provide coordination of functional activities of the smooth muscles of intestinal wall to make propagation of the content less depending on the mechanical factors like the wall resistance and semi liquid consistency of the content.

The second equation in (*) determining changes in a muscle tone $\frac{dN}{dt} = F(N, \varepsilon)$ should be modified so that to reflect both possibilities. In this case $\frac{dN}{dt} = F(N, \varepsilon, f)$ and equation considering the influence of external neuronal factor on changes in the muscle tone has a view

$$\frac{dN}{dt} = -k_3 N^3 + k_2 N^2 - k_1 N + \varkappa \varepsilon + \alpha f \quad (2)$$

For simplicity $k_0(t)$ is omitted. Expression for $f$ includes parameter $\alpha$ defined as the pressure differential between the current pressure inside the reservoir $p$ and the threshold pressure when the output valve opens $\pi_-$ :

$$a = \frac{a^*(p - \pi_-)}{\pi_-} \quad (3)$$

Here $a^*$ is a positive coefficient referring the pressure differential to $f$ .

In this case, like in [1] pressure inside the reservoir $p$ is a linear function of two variables $p = L(\varepsilon, N)$, though smooth muscle tone $N$ is also affected by external activation factor $f$ .

In order to obtain closed system showing influence of smooth muscle tone $N$ on neuronal activator $f$ conditions determining the changes of neuronal activator $f$ should be formulated based on physiology of intestinal motility. According to the existing data changes in the concentration of neuronal activator should be affected by the stressed conditions of the smooth muscles in a manner that the increase in the muscle tone during filling should inhibit the flow of the external activator into the extracellular space during filling and increase it during emptying, to contribute to the muscles contractions.

It is logical to assume that baroreceptors located in the intestinal wall are used by peripheral neural systems (Meissner's and Auerbach's plexuses) to obtain information about stress-strained conditions of the smooth muscles in order to send inhibitory or excitatory stimuli back to the muscles.

Modified equation (1) considering the influence of muscle tone on external activator has the form:

$$\frac{df}{dt} = -l_3 f^3 + l_2 f^2 - l_1 f + bN \quad (4)$$

Parameter $b$ like parameter $a$ in (3) is a function of the pressure inside the reservoir,

$$b = \frac{b^*(p - \pi_-)}{\pi_-} \qquad (5)$$

$b^*$ - coefficient relating $b$ to $N$.

It is easy to see that $a$ and $b$ in (3) and (5) may have positive or negative values depending on the pressure differentials.
Negative values of the parameter $b$ in (5) indicate inhibitory actions of the muscle tone on external activator during filling when pressure inside the reservoir is below $\pi_-$. It has a physiologic sense which is preventing overstimulation of the muscles in order to keep them compliant while the filling continues. The same meaning has the parameter $\alpha$ in (2). Until the pressure reaches the level when input valve closes $\alpha$ has negative values preventing quick increase of the muscle tone, rise of the intraluminal pressure and early emptying.

During isovolumetric states the pressure inside reservoir is between $\pi_+$ (input valve closes) and $\pi_-$, giving a negative sign to $b$ and positive to $a$. This will correspond to the NFB loops between internal and external activators. During emptying phases the pressure inside the reservoir $p$ exceeds $\pi_-$, so that parameters $\alpha$ and $b$ have positive sign. It results in simultaneous activation of myogenic and neurogenic components of the muscle tone through PFB mechanisms. In other words, it makes myogenic and neurogenic regulatory mechanisms activate one another which results in a quick rise of the pressure and increased propulsive forces while emptying. Transition from isovolumetric to the emptying phase will correspond to the bifurcation of phase portrait due to topological incompatibility of NFB and PFB dynamical images.

All possibilities are clearly demonstrated by the view of the matrices $A_{lin}$ of a linearized system (3) and (4) around some equilibrium point $(N^*, f^*)$ during filling, isovolumetric stage and emptying:

$$\frac{dN}{dt} = -k_3 N^3 + k_2 N^2 - k_1 N + af$$

$$\frac{df}{dt} = -l_3 f^3 + l_2 f^2 - l_1 f + bN$$

$$A_{lin} = \begin{pmatrix} \pm k & \pm a \\ \pm b & \pm l \end{pmatrix} = \begin{pmatrix} \pm\alpha(N) & \pm\beta(f) \\ \pm\gamma(N) & \pm\delta(f) \end{pmatrix} \qquad (6).$$

Both $\beta$ and $\gamma$ coefficients are negative during filling, have opposite signs while the muscle tone is experiencing isovolumetric changes, and are positive during the emptying. Depending on the values of the coefficients, making sense scenarios and related to them phase portraits are: (i) saddle, if determinant $A_{lin} < 0$ and $\alpha\delta < \beta\gamma$, (ii) steady node if det $A_{lin} < 0$, $\alpha\delta > \beta\gamma$ and (iii) steady focus, if det $A_{lin} > 0$, $\alpha\delta < \beta\gamma$. In the last case skew-diagonal entries have opposite signs ($\beta\gamma < 0$). Like in a previous work [1], where the basis functional pattern, NFB, was not shown explicitly in the system (*), $A_{lin}$ also can be presented as a sum of a diagonal and a skew diagonal matrices, where skew diagonal matrices will represent PFB or NFB patterns (Fig. 3).

In case of the saddle phase trajectories show simultaneous increase or decrease values of $N$ and $f$ variables on a physiologically permissible area where the values of the variables are positive. It may indicate mutual activation or inhibition of internal and external activation factors by PFB loops.

On the planes of $(N_1, \varepsilon_1)$ and $(N_2, \varepsilon_2)$ variables $N$ and $\varepsilon$ from [1] are split between two connected reservoirs each in different motility phases: $(N_1, \varepsilon_1)$- for emptying reservoir (I), and $(N_2, \varepsilon_2)$ for reservoir (II) being filled (Fig. 6). All assumptions of the first model (*) remain the same with some additional requirements related to the necessity to coordinate the reservoirs emptying-filling functions.

In the first equation (*) constant values of input ($P_+$) and output ($P_-$) threshold pressures must be replaced by the corresponding current pressures $p_1$ and $p_2$ inside the reservoirs (I) and (II) considered as functions of stretching (deformations) $\varepsilon$ and muscle tone $N$.

$$p_1 = \frac{(E\varepsilon_1 + N_1)h_0}{(2\varepsilon_1 + 1)R_0} \quad (7)$$

$$p_2 = \frac{(E\varepsilon_2 + N_2)h_0}{(2\varepsilon_2 + 1)R_0} \quad (8)$$

For simplicity some of the coefficients are omitted.
The coordinated wall motions of sequentially connected to one another two distensible cylindrical reservoirs can be described by a system of six ODEs of three types of variables, each type for two reservoirs: stretching ($\varepsilon_{1,2}$), smooth muscle tone ($N_{1,2}$) and external activator ($f_{1,2}$). Indices correspond to the proximal and distal reservoirs.

$$\frac{d\varepsilon_1}{dt} = \frac{1}{2V_0}\left(\frac{P_+}{Z_+} + \frac{p_2}{Z_-} - \frac{E\varepsilon_1 + N_1}{2\varepsilon_1 + 1}\frac{h_0}{R_0}\frac{1}{Z}\right) \quad (9)$$

$$\frac{d\varepsilon_2}{dt} = \frac{1}{2V_0}\left(\frac{p_1}{Z_+} + \frac{P_-}{Z_-} - \frac{E\varepsilon_2 + N_2}{2\varepsilon_2 + 1}\frac{h_0}{R_0}\frac{1}{Z}\right) \quad (10)$$

$$\frac{dN_1}{dt} = -k_{31}N_1^3 + k_{21}N_1^2 - k_{11}N_1 + \varkappa_1\varepsilon_1 + a_1f_1 \quad (11)$$

$$\frac{dN_2}{dt} = -k_{32}N_2^3 + k_{22}N_2^2 - k_{12}N_2 + \varkappa_2\varepsilon_2 - a_2 f_2 \qquad (12)$$

$$\frac{df_1}{dt} = -l_{31}f_1^3 + l_{21}f_1^2 - l_{11}f_1 + b_1 N_1 \qquad (13)$$

$$\frac{df_2}{dt} = -l_{32}f_2^3 + l_{22}f_2^2 - l_{12}f_2 - b_2 N_2 \qquad (14)$$

The first two equations (9) and (10) are based, like (1), on the mass conservation law with separate pressure inside the reservoirs depending on the current values of wall deformations and smooth muscle tones. The next two equations (11), (12) describe changes in muscle tones of each reservoir affected by intrinsic ($N$) and extrinsic ($f$) activation factors. The last two equations (13), (14) relate changes in external activation factors depending on inner neuronal mechanisms and muscle tones.

1.2  REGULATIONS OF WALL MOTIONS OF TWO CONNECTED RESERVOIRS

Comparison of ($N_1$, $\varepsilon_1$) and ($N_2$, $\varepsilon_2$) dynamical images shows that isoclines $\frac{dN_1}{dt} = 0$, $\frac{dN_2}{dt} = 0$ corresponding to the phase trajectories for reservoir (I) and reservoir (II) are moving simultaneously in opposite directions relative to the $\varepsilon$ axes. It changes the slopes of the phase curves, demonstrating the slowing down both filling and emptying phases. Simultaneous activation of muscle tone of reservoir (I) (emptying) and inhibition of smooth muscles contractility of reservoir (II) (filling) improve functional compliance of intestinal wall resulted in alleviation of the passage of the content from one intestinal segment to another (Fig. 7, 8).

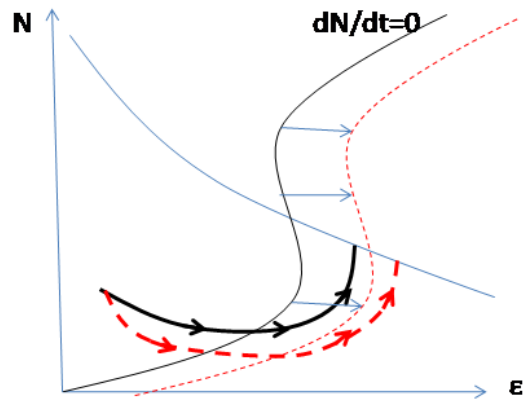

**Figure 7  Influence of external activator on the character of the filling curve**

Displacement of isocline dN/dt=0 shows improvement in smooth muscles compliance resulted in increase of the filling volume of reservoir (dashed curve). Isovolumetric area and portions of the curves are not shown on the sketch.

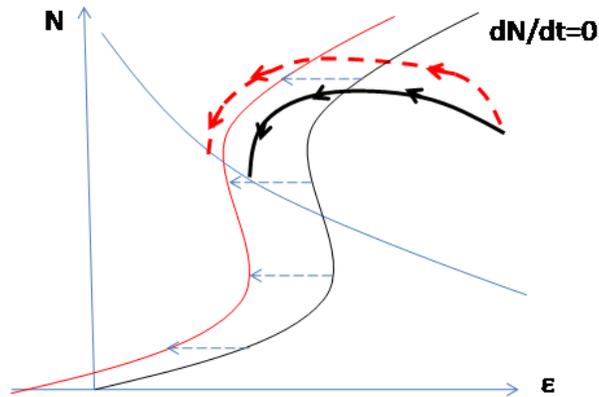

Figure 8 Phase curves showing influence of the external activator on emptying

Displacement of isocline dN/dt=0 shows improvement in contractility of the reservoir's wall resulted in increase of evacuating volume (dashed curve). Isovolumetric area is not shown on the sketch.

Comparison of wall movements of two reservoirs also indicates possibility for reciprocal relationships. It is easy to show that on the plane of $\varepsilon_1$ and $\varepsilon_2$ variables, (isoclines $\frac{d\varepsilon_1}{dt} = 0$ and $\frac{d\varepsilon_2}{dt} = 0$ are considered monotonous functions), phase portrait around the equilibrium can be saddle or steady node (Fig. 9).

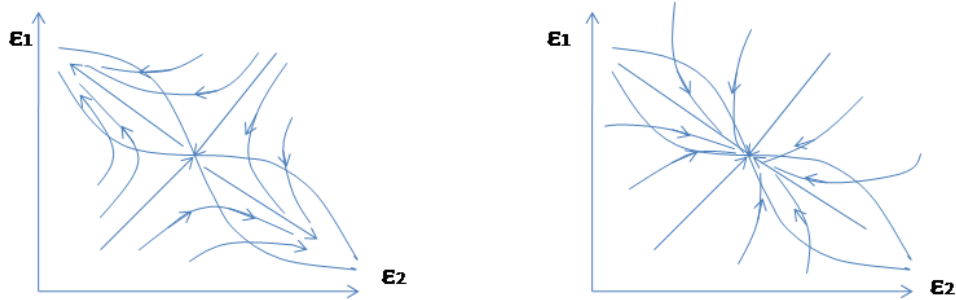

Figure 9 Saddle (left) and steady node (right) are two types of functional patterns coordinating wall movements of two reservoirs. Left image (saddle) indicate reciprocal relationships between $\varepsilon_1$ and $\varepsilon_2$

The obtained results demonstrate a physiologic advantage of including of external activator in motility regulations. It also confirms quite obvious fact that phylogenetically developing mechanism of propagation of the content along GI tract requires reciprocal functional interactions of neighbor intestinal segments.

Described above relationships can be summarized by a diagram where each group of variables $N_1, f_1, \varepsilon_1$ and $N_2, f_2, \varepsilon_2$ related to the connected to one another reservoirs includes NFB and possibly PFB, while myogenic activation, neuronal activation mechanisms and stretching of both reservoirs are linked by RL or PFB regulatory patterns (Fig. 10).

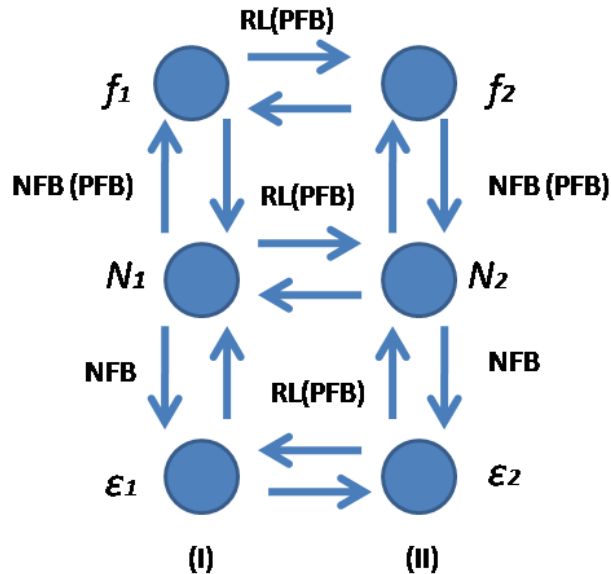

**Figure 10** Schematic representation of basis functional patterns between $N, \varepsilon, f$ variables considered separately for two connected reservoirs (I) and (II).
RL and topologically equivalent PBF patterns (in parentheses) relate same types of variables of two reservoirs. Patterns, responsible for steady nodes are not shown; the sketch represent only imaginary basis elements of coquaternion.

DISCUSSION

GI motility can be understood from two different approaches to physiologic mechanisms: the first one is related to the functional mechanisms regulating myogenic active properties of smooth muscles of intestinal wall; the second one is related to the functional patterns collaborated in a functional system regulating motility of GI segments.
Three basis functional elements known as stable, reproducible patterns regulating functions on different levels including organs, biological cells and biochemical reactions are NFB, PFB and RL (PNR). With the identity element they can be presented in the form of the basis elements of the algebraically closed structure, - a coquaternion, which is a ring of matrices. The imaginary part of coquaternion consists of three basis elements having matrix structures of NFB, PFB and RL. The fact that any invertible 2x2 matrix over real numbers can be presented as a combination of coquaternion basis elements is a reason to consider a non-singular functional regulatory structure of a biologic system expressed in a matrix form as a combination (integration) of its basis functional elements.

NFB is shown as a main functional pattern regulating stress-strained conditions of smooth muscles of intestinal wall. Its regulatory structure contributes to the steady focus whose curves demonstrate changes of the stress-strained conditions of intestinal wall from some initial state to the equilibrium point. Trajectories of steady focus divide motility process into three phases:

filling, isovolumetric increase of reservoir's pressure and emptying. Mechanisms of regulations of the trans-membrane flow of the myogenic factor (activator), which is ionized calcium $Ca^{2+}$, was used as a physiologic background for formulating dynamical equation for the muscle tone $N$ in the previous model (*) [1].

If emptying shown on the phase portrait of ε and $N$ variables of (*) is considered for the proximal of two attached reservoirs and trajectories corresponding to filling are related to the distal reservoir, then according to normal physiology and in a very simplified modeling conditions movement of the content from one portion of intestine to another should be provided not only by myogenic active properties of the smooth muscles of intestinal wall, but also through the additional, neurogenic, mechanism alleviating emptying and propagation of the content. External, neuronal, activator supposed to slow down the process of diminishing muscle tone of the reservoir being emptied in order to keep the contractile forces in the action for longer time. At the same time, external activator should further decrease muscle tone of the reservoir being filled before meeting the isocline $\frac{dN_2}{dt} = 0$ and after the intersection partially inhibit an increase in the smooth muscle tone due to its activation caused by the wall distension. It shows that physiologic advantages of the demonstrated mechanism of coordination of the motions of intestinal walls of two reservoirs regulated by external activator are provided by reciprocal interactions of proximal and distal segments. It finds its explanation in the fact that during phylogenetic development morphological differentiation occurs simultaneously with diversity of their functions linked and realized by reciprocal mechanisms.

Reciprocal interactions were also explicitly shown between internal, regulating the flow of ionized $Ca^{2+}$, and external, regulating the transmembrane flow of $Na^+$, activation mechanisms altogether regulating contractility of the smooth muscles of an isolated intestinal segment. Reciprocal interactions between activation mechanisms of proximal and distal reservoirs were also demonstrated as an adjustment of evacuatory efforts of the emptying reservoir to the muscles compliance of the reservoir being filled, which is confirmed by opposite displacements of $\frac{dN_1}{dt} = 0$, $\frac{dN_2}{dt} = 0$ isoclines corresponding to the emptying and filling phases.

Images of PFB and RL related flows are topologically equivalent as saddles. It means that linear combinations of chosen variables allow one to demonstrate that these two functional structures represent equivalent regulatory patterns relative to the new basis variables obtained as linear combinations of the old ones. In general, a matrix describing the system's behavior can be presented as a sum of a skew diagonal one representing PFB and a diagonal matrix related to steady or unsteady node images.

Because of similar cubic expressions for $N$ (internal activator) and $f$ (external activator) having negative as well as a positive signs of the parameters, there is a possibility for different signs of the entries of diagonal elements of linearized systems depending on the parameters values. This is possible if domain corresponding to the activation of one variable will coincide with the domain related to the depression of another. It will indicate the existence of the RL patterns (or regulatory component related to RL) for these variables because of opposite signs of diagonal elements. This diagonal matrix most probably will be a summand of a skew diagonal component. Therefore, motility mechanisms of intestinal segments suppose to include all three (PNR) functional patterns responsible for the regulatory function of intestine.

Matrices of PNR have the form of imaginary basis elements of coquaternion, which with the matrix determining steady (unsteady) node provide the model simulating intestinal motility with four basis elements comprising algebraic structure of coquaternion [21, 22, 27, 33, 34]. A set of coquaternions is closed as a ring of matrices under addition and multiplication of its elements. Hypothetically, GI tract or its segment (portion of intestine) can be considered as an autoregulatory anatomical unit. Physiologically it means that integration of three basis functional patterns (PNR) including a pattern associates with the scalar (in steady/unsteady node) will provide the system with adequate outcome and maintain wholeness of its internal functional structure.

ACKNOLEDGMENT

I would like to thank Wei Wei Wang, Janetta Kourzenkova M.D. and Hilda Berghuis for discussions and technical assistance while preparing the paper

CONFLICT OF INTERESTS

This work was not supported by any funds. No conflict of interests is declared.